\documentclass[11pt]{JHEP3}
\usepackage{graphicx,amssymb}
\usepackage{epsfig}

\newcommand{\newc}{\newcommand}
\newc{\eeq}{\end{equation}}
\newc{\beq}{\begin{equation}}
\newc{\ec}{\end{center}}
\newc{\bc}{\begin{center}}
\newc{\eeqa}{\end{eqnarray}}
\newc{\beqa}{\begin{eqnarray}}
\newc{\nonr}{\nonumber}
\newc{\PD}{\partial}
\newc{\bd}{\begin{description}}
\newc{\ed}{\end{description}}
\newc{\benu}{\begin{enumerate}}
\newc{\eenu}{\end{enumerate}}
\newc{\bi}{\begin{itemize}}
\newc{\ei}{\end{itemize}}
\newc{\ra}{\rightarrow}
\newc{\bis}{\begin{itemstep}}
\newc{\eis}{\end{itemstep}}
\newc{\lag}{\cal{L}}
\newc{\LH}{\hat{L}}
\newc{\RH}{\hat{R}}
\newc{\SL}{\not\!}
\newc{\mtr}{\mathrm {tr}}
\newc{\ol}{\overline}
\newc{\Gf}{\frac{G_F}{\sqrt 2}}

\newc{\ugl}{\ol{u_L}^i\gamma^{\mu}u_{L}^i}
\newc{\ugr}{\ol{u_R}^i\gamma^{\mu}u_{R}^i}
\newc{\egl}{\ol{e}^k\gamma_{\mu}\LH e^k}
\newc{\egr}{\ol{e}^k\gamma_{\mu}\RH e^k}
\newc{\dgl}{\ol{d}^i\gamma^{\mu}\LHd^i}
\newc{\dgr}{\ol{d}^i\gamma^{\mu}\RHd^i}
\newc{\alo}{\alpha_1}
\newc{\alt}{\alpha_2}
\newc{\alth}{\alpha_3}
\renewcommand{\theenumi}{\Alph{enumi}}

\title{An Effective Operators Analysis of CP Violation : The Semileptonic Case }
\author{We-Fu Chang\\Institute of Physics, Academia Sinica, Nankang, Taipei, Taiwan 115\\
E-mail: \email{wfchang@phys.sinica.edu.tw}}
\author{John N. Ng\\ Theory Group,
TRIUMF, 4004 Wesbrook Mall,
Vancouver, B.C.,
Canada V6T 2A3\\ E-mail: \email{misery@triumf.ca}}

\abstract{
Aiming at a model-independent analysis of possible  new physics effects
in semileptonic processes at various energy scales,
we list and study a complete set of $SU(3)_c\times SU(2)_L\times U(1)_Y$ invariant
4-Fermi operators which consist of a pair of quarks and a pair of leptons above the electroweak
symmetry breaking.
We give a full 1-loop  renormalization group treatment of the evolution of the Wilson coefficients
associated with these 4-Fermi operators between low energy ($\sim$ meson masses)
and the cutoff scale $\Lambda$, $\sim (1-10)$ TeV, where
we assume  new degree of freedom beyond standard model will begin to appear and an ultra-violet
completion of our effective theory will take place.

Motivated by the existing phenomenological bounds, we argue that the new CP
violation can only stem from the scalar and tensor types of  4-Fermi
interaction.
Some interesting constraints are obtained  by studying  the universality of kaon
and pion leptonic decays, CP violating polarization of $K^+_{\mu 3}$,
charged lepton anomalous magnetic moments, and $(\mu\ra e \gamma)$ like rare decays.
In particular, we can use the limit  of electron dipole moment
to constrain the size of the CP violating triplet correlation in the $e^+ e^- \ra t \bar{t}$
process.
}

\begin{document}

\section{Introduction}

Intense experimental efforts in the search for new physics have  resulted in establishing the
Standard Model (SM) as a  very good effective theory at the weak scale given by
the Higgs boson vacuum expectation value of $v\simeq  250$ GeV and below. Yet
there are strong arguments that it is one with a cutoff scale $\Lambda$
much lower than the Planck scale; perhaps
as low as a few TeV. More importantly the observation of neutrino mixings have given the first hint
new physics beyond $v$. The exact form of physics beyond the SM is unknown. One can proceed by
making a guess at this new physics and engage in constructing consistent models.
The other approach is to make
use of the effective field theory (EFT). The main assumption here is that at energies below $\Lambda$
physical observables are largely insensitive to the unknown new physics. The effective
Lagrangian is then a sum of the SM term and non-renormalizable ones which are the results
of integrating out the unknown degrees of freedom. This bottom up approach clearly has its
drawbacks. One of them being that there can be many such non-renormalizable terms. This is not as
hopeless as it seems. This operators can be classified by their dimensions and the higher
ones are suppressed by higher powers of $1/\Lambda$. Strictly speaking the cutoff scale for
each set of operators need not be the same. Hence, we need only focus on the lowest
dimension operators.

The EFT approach is built upon two crucial ingredients.
The first one to correctly identify the symmetry operative below the scale $\Lambda$. The
second one is to know  the degrees of freedom.
 The phenomenal success of the SM in confronting experiments suggest strongly that
 we take the gauge symmetry of the SM to be operative between $\Lambda$ and $v$. Below
the electroweak scale the gauge symmetry is $SU(3)\times U(1)$. To this
 we may also incorporate baryon number so as to ensure proton stability. This is not
mandatory as is well known that operators leading to proton decay can be suppressed by
mass scale of order $10^{16}$ GeV. While taking the SM as the gauge symmetry below $\Lambda$ is
largely not controversial the same cannot be said about what states to include in constructing an EFT
below $\Lambda$. The most conservative paradigm is to take only the SM fields of 45 chiral fermions
plus the gauge bosons and one Higgs doublet. The effective Lagrangian will then take the form
\beq
\label{Leff}
{\lag}^{eff}= {\lag}_{\mathrm {SM}} +\frac{1}{\Lambda^{\prime}}{\lag}_5 + \frac{1}{\Lambda^2}{\lag}_6 + \cdots.
\eeq
Here ${\lag}_5$ is the dimension 5 operator constructed from the neutrino and Higgs fields
which is responsible for generating Majorana neutrino masses for the active neutrinos. The seesaw scale
is denoted by $\Lambda^{\prime}$ which may or may not be the same as  $\Lambda$ depending on the origin of
neutrino masses.
If this
is non-vanishing then lepton number is not conserved. On the other hand neutrinos may be Dirac particle
in which case ${\lag}_5$ vanished\footnote[2]{ It is sufficient to have a $U(1)_{B-L}$ symmetry to ensure proton stability
and no Majorana neutrino mass. However,we are mindful that
Planck scale physics is likely to break such global symmetry.}.
The details of how neutrinos get their masses will not
be important and we only note that active neutrinos are massive and they mix.
${\lag}_6$ is a sum of dimension 6 operators composed of chiral fermions, gauge bosons,
 and the Higgs field. The dots
are higher dimension operators suppressed by higher powers of $1/\Lambda$. The number of operators
in ${\lag}_6$ is over 20 \cite{oplist} even after the use of equations of motion and not counting
family dependence. Clearly  a  purely  phenomenological analysis will be unwieldy. It will be a more tractable problem if we select a subset of these that are closed under renormalization
to 1-loop and analyze their physical effects.

We begin with the  dimension 6 four-Fermi operators that are made up of a pair
of leptons and a pair of quarks.
We are especially interested in  the ones give CP violating (CPV) effects\footnote{
Since we are considering only semileptonic decays and lepton dipole moments
the 4-quark operators will  not be included here.}.
There are 10 such operators which we
will  divide into two groups: the vector type, and the  scalar and tensor type. The vector terms are
explicitly,
\beqa
\label{OPV}
O_{V1}^{ij,kl}&=&(\bar{Q}^i \gamma^{\mu}Q^j)(\bar{L}^k \gamma_{\mu}L^l)\,,\\
O_{V2}^{ij,kl}&=&(\bar{Q_a}^i\gamma^{\mu}{Q_b}^j)(\bar{L_b}^k\gamma_{\mu}{L_a}^k)\,,\\
O_{V3}^{ij,kl}&=&(\bar{Q}^i\gamma^{\mu}Q^j)(\bar{e}^k\gamma_{\mu}e^l)\,,\\
O_{V4}^{ij,kl}&=&(\bar{d}^i\gamma^{\mu}d^j)(\bar{L}^k\gamma_{\mu}L^l)\,,\\
O_{V5}^{ij,kl}&=& (\bar{u}^i\gamma^{\mu}u^j)(\bar{L}^k\gamma_{\mu}L^l)\,,\\
O_{V6}^{ij,kl}&=&(\bar{d}^i\gamma^{\mu}d^j)(\bar{e}^k\gamma_{\mu}e^l)\,,\\
O_{V7}^{ij,kl}&=&(\bar{u}^i\gamma^{\mu}u^j)(\bar{e}^k\gamma_{\mu}e^l)\,,
\eeqa
and the remaining scalar and tensor terms are
\beqa
\label{OPST}
O_{S1}^{ij,kl}&=&(\bar{Q}^i d^j)(\bar{e}^k L^l)\,,\\
O_{S2}^{ij,kl}&=&(\bar{Q_a}^i u^j)(\bar{L_b}^k e^l)\epsilon^{ab}\,,\\
O_T^{ij,kl}&=&(\bar{Q_a}^i \sigma^{\mu \nu} u^j)(\bar{L_b}^k\sigma_{\mu \nu}e^l)\epsilon^{ab}\,,
\eeqa
where $Q,L,e,d,u$ are respectively the left-handed quark doublets, left-handed lepton doublets,
right-handed charged leptons, down and up-type quarks of the SM. $a,b$ are $SU(2)$ indices
and $i,j,k,l$ are family indices.
We have also suppressed the chiral projection operators.
We use the convention that repeated indices are summed over and our  metric $g^{\mu\nu}$ is $(+---)$
with $\{\gamma^{\mu},\gamma^{\nu}\}=2g^{\mu\nu}$. The purely leptonic 4-Fermi
operators are recently studied in detail in \cite{CN1} and we shall follow the notations given there.
Then in the weak basis we write
\beq
\label{L6}
-{\lag}_6=\sum_{A=1}^7 C_{VA}^{ij,kl}O_{VA}^{ij,kl} +\sum_{A=1}^2 C_{SA}^{ij,kl}O_{SA}^{ij,kl}
+C_T^{ij,kl}O_T^{ij,kl} +h.c.
\eeq
In general the Wilson coefficients denoted by $C$'s depend on the family index.
Also hermiticity requires the 7 vector Wilson coefficients to
satisfy
\beq
C_V^{jilk}= \left( C_V^{ijkl} \right)^*\,.
\eeq
However, for the scalar and tensor types there is no such
requirement. Clearly the most general flavor structure is too unwieldy for a fruitful analysis and
we shall make phenomenologically motivated simplification that hopefully reflects some dynamics
of the yet unknown new physics.

Even after going over to the mass basis of the fermions there will be
flavor changing neutral currents (FCNC) generated at 1-loop. These are  known to be highly suppressed.
Following the discussions in \cite{CN1} we make the ansatz
\beq
C_{VA}^{ij,kl}=C_{VA}^{ii,kk}\delta^{ij}\delta^{kl}\,,
\eeq
 where $C_V^{ii,kk}$ are constants that depend on the quark and lepton families.
The operator $O_{V2}$ contains charged weak currents.
Universality tests of tau and Pion decays suggest that their Wilson coefficients should
be independent of lepton flavors. Thus it is reasonable to assume
\beq
C_{V2}^{ii,kk}=C_{V2}^{ii}\,.
\eeq
We shall see in the next section that renormalization will mix $O_{V1}$ and $O_{V2}$ and hence to
satisfy the above mentioned experimental constraints we take $C_{V1}^{ii,kk}=C_{V1}^{ii}$.
Thus, we are emboldened to extend it  to all vector Wilson coefficients  and drop the dependence
on the families indices. In fact most four dimensional unified models have this feature.
Some notable exceptions are provided by extra dimension models and split fermion models \cite{CN2}.
With that in mind we rewrite the effective Lagrangian in the mass eigenbasis as
\beq
\label{L6mass}
-{\lag}_6=\sum_{A=1}^7 C_{VA}O_{VA}^{ii,kk}+\sum_{A=1}^2C_{SA}^{ij,kl}O_{SA}^{ij,kl}+ C_T^{ij,kl}O_T^{ij,kl}+h.c.
\eeq

From now on all our  operators expressed will be in the  mass eigenbasis. Notice that the flavor
insensitivity  of the vector operators are not extended to the scalar and tensor operators.
Certainly the $C_{S,T}$'s in Eq.(\ref{L6mass}) differ from those in Eq.(\ref{L6}) by products
of rotational matrices that diagonalize the fermion masses.
For simplicity we have kept the same notation.

There are eight dimension-6 dipole operators with two fermions. They are listed as
\beqa
\label{eq:dipole1}
D_1^{ij} &=& H \bar{L}^i \sigma^{\mu\nu} e^j B_{\mu\nu}\,,\\
\label{eq:dipole2}
D_2^{ij} &=& H \bar{L}^i \sigma^{\mu\nu} e^j W_{\mu\nu}\,,\\
D_3^{ij} &=& \widetilde{H} \bar{Q}^i \sigma^{\mu\nu} u^j B_{\mu\nu}\,,\\
D_4^{ij} &=& \widetilde{H} \bar{Q}^i \sigma^{\mu\nu} u^j W_{\mu\nu}\,,\\
D_5^{ij} &=& \widetilde{H} \bar{Q}^i \sigma^{\mu\nu} u^j G_{\mu\nu}\,,\\
D_6^{ij} &=& H \bar{Q}^i \sigma^{\mu\nu} d^j B_{\mu\nu}\,,\\
D_7^{ij} &=& H \bar{Q}^i \sigma^{\mu\nu} d^j W_{\mu\nu}\,,\\
D_8^{ij} &=& H \bar{Q}^i \sigma^{\mu\nu} d^j G_{\mu\nu}\,,
\eeqa
where $\tilde{H}= i\sigma^2 H^*$, and $B, W, G$ are the field
strength of $U(1)_Y, SU(2)_L$ and $SU(3)$ respectively.

Notice that the operators are in general not flavor diagonal. After electroweak  symmetry breaking (EWSB)
we can substitute $H\ra (v+h^0)/\sqrt 2$. The  terms from Eq.(\ref{eq:dipole1},\ref{eq:dipole2}) become
\beq
-{\lag}_6^{D}= \frac{v C_D^{ij}}{\sqrt 2}(\ol{e}^i\sigma^{\mu\nu}\RH e^j F_{\mu\nu})+h.c.
\eeq
They will induce
flavor violating or conserving radiative transitions that result in
lepton electric dipole moments (EDM's) , $(g-2)_\mu$, $\mu\ra e \gamma$,
 and $b \ra s \gamma$  at the level of the Born term.  All these  are strongly
constrained by null experimental observation beyond SM \cite{CN1}. For example, the muon $g-2$
constrains ${\frak{Re}}\, C_D^{22}$ and electron EDM limits ${\frak{Im}}\,C_D^{11}$
and they have to be $<10^{-6}$.
>From this  observation we shall adopt a phenomenologically
motivated assumption that any reasonable new physics beyond SM
will only induce negligible tree-level dipole operators.
\begin{figure}[h]
      \begin{center}
    \epsfig{file=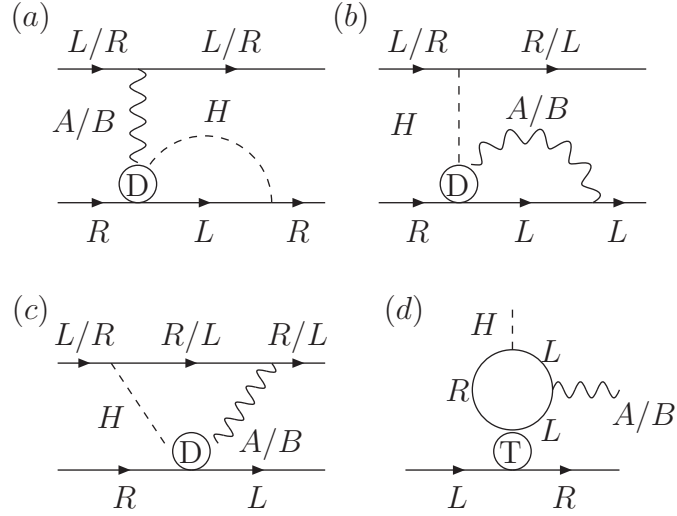, width=0.6\textwidth}
    \end{center}
  \caption{The 1-loop diagrams which mix up the dipole and 4-Fermi operators.
  Some representative chirality configuration are shown. $A/B$ stand for
  the $SU(2)_L/U(1)_Y$ gauge boson. }
\label{fig:opmix4D}
\end{figure}

Proceeding further one recognizes that the dipole operators can be
induced at 1-loop by the 4-Fermi operators and SM interactions.
In  Fig.\ref{fig:opmix4D} we show the possible 1-loop
diagrams that will mix the 4-Fermi operators and the dipole
operators. The first three terms are vanishing from the above discussion. Since
only the t-quark enjoys ${\cal O}(1)$
Yukawa coupling and thus the  mixing involving other light quarks are severely
suppressed by their small Yukawa couplings and can be neglected. Fig.\ref{fig:opmix4D}(d) shows that
 we only need to consider the
 dipole term induced from the tensor operator involving the third
generation quarks, $O_T^{33,kl}$.
Thus, we only need to consider the first two dipole operators
involving 2 leptons.
 At low energies for EDM studies the relevant operator involves the photon thus we define
\beq
\label{dipole2}
O_{D\gamma}^{ij} = \bar{L}^i\sigma^{\mu\nu}e_R^j F_{\mu\nu}H +h.c.
\eeq
and the Wilson coefficient is $C_{D\gamma}^{ij}= c_{d1}^{ij}\cos \theta_w +c_{d2}^{ij}\sin\theta_w$ in standard notations. The complete
effective Lagrangian at the weak scale is now
\beq
\label{LEL}
-{\lag}_6=\sum_{A=1}^7 C_{VA}O_{VA}^{ii,kk}+\sum_{A=1}^2C_{SA}^{ij,kl}O_{SA}^{ij,kl}+ C_T^{ij,kl}O_T^{ij,kl}
+\sum_{A=1}^2C_{DA}^{ij} O_{DA}^{ij} +h.c.
\eeq

The physics is clearer if we expand ${\lag}_6$  in terms of their components.
We split it into the  charged current (CC) terms and the neutral current (NC) terms.
After some algebra the  NC terms are given by,
($\LH=\frac{1-\gamma_5}{2},\RH=\frac{1+\gamma_5}{2}$),
\beqa
\label{LNC}
-{\lag}_6^{\mathrm{NC}}&=&
\ol{u}^i \gamma^{\mu} \left[ (C_{V1}+C_{V2})\LH
++C_{V5}\RH\right] u^i \times (\ol{\nu}^k \gamma_{\mu}\LH\nu^k) \nonr \\
&+& \ol{d}^i \gamma^{\mu}  \left[C_{V1}\LH +C_{V4}\RH\right]d^i\times(\ol{\nu}^k\gamma_{\mu}\LH\nu^k) \nonr \\
&+&\ol{u}^i\gamma^\mu \left[ C_{V1}\LH + C_{V5}\RH \right]u^i\times (\egl) \nonr \\
&+&\ol{d}^i\gamma^\mu \left[(C_{V1}+C_{V2})\LH + C_{V4}\RH \right]d^i\times (\egl) \nonr \\
&+&\ol{u}^i\gamma^\mu \left[ C_{V3}\LH + C_{V7}\RH \right]u^i\times (\egr) \nonr \\
&+&\ol{d}^i\gamma^\mu \left[ C_{V3}\LH + C_{V6}\RH \right]d^i\times (\egr) \nonr \\
&+& C_{S1}^{ij,kl}(\ol{d}^i\RH  d^j)(\ol{e}^k \LH e^l)+ C_{S2}^{ij,kl}(\ol{u}^i \RH u^j)(\ol{e}^k\RH e^l)
\nonr \\
&+&C_T^{ij,kl}(\ol{u}^i\sigma^{\mu \nu} \RH u^j)(\ol{e}^k\sigma_{\mu \nu} \RH e^l) +h.c.
\eeqa

If one probes the effective neutral current couplings using neutrinos versus electrons one would find the corrections to the SM
are very different.  Interestingly the NC effects of the electron can contain terms with different Lorentz structure
then that of the SM. On the other hand the new NC couplings of the neutrinos are  only of the vector and axial
vector types. This follows from the assumption that there are no right-handed neutrinos below $\Lambda$ and the gauge
symmetry of the SM. The structure of Eq.(\ref{LNC}) also points to polarized electrons being  powerful and versatile
probes of new physics.

The CC terms are explicitly:
\beqa
\label{LCC}
-{\lag}_6^{\mathrm {CC}}&=& C_{V2}^{\prime}(\ol{u}^i\gamma^{\mu}\LH d^i)(\ol{L}^j\gamma_{\mu}\LH \nu
^i)+C_{S1}^{ij,kl}(\ol{u}^i\RH d^j)( \ol{e}^k \LH \nu^l) \nonr \\
&& - C_{S2}^{ij,kl}(\ol{d}^i \RH u^j)(\ol{\nu}^k \RH e^l) -C_T^{ij,kl}(\ol{d}^i\sigma^{\mu \nu} \RH u^j)(\ol{\nu}^k\sigma_{\mu \nu} \RH e^l) \nonr \\
&& +h.c.
\eeqa
$C_{V2}^{\prime}$ differs from its NC counterpart by quark and lepton rotational matrices.
In principle if the parameters involved are all measured in  the future the relation between
$C_{V2}$ and $C_{V2}^{\prime}$ will be a test the above framework.
Currently we only have limits on these parameters.
The Eq.(\ref{LCC}) renormalizes the canonical 4-Fermi interaction; thus $C_{V2}^{\prime}$
can be revealed by precision measurements comparing tau and muon leptonic decays with
neutron beta decays as done in classic universality tests.
The scalar and tensor terms adds incoherently and hence are not important.
Nonetheless, these latter terms are very interesting since their discovery will be strong
indication of new physics.
They  can be probed by $K$ and $B$ meson decays;  for example by measuring the energy spectral
of the charged leptons in semileptonic decays.
This is  analogous to what we have studied previously  in $\mu$ decays \cite{CN1}.
Furthermore, the Wilson coefficients  are in general complex and thus can give rise
to novel CP violation effects via interference with a SM amplitude.
A prime example will be  muon polarization measurements in
$K^+\ra \pi^0\mu^+ \nu$ which are unique low energy probes of these terms \cite{Kl3}.
Details  will be given later.

Now we return  to more theoretical issues.
The Wilson coefficients are  defined at the scale $\Lambda$.
In a top down approach one would be able to calculate them in terms of the
parameters of the new physics theory.
The bottom up path taken here  does not afford such luxury.
We can only treat them as parameters to be determined or constrained  by experiments.
To date all relevant CPV experiments are done at energies much lower than $\Lambda$.
We must then evolve the Wilson coefficients from $\Lambda$ to $v$ and then to the
scale of $K$, $B$ mesons or charged lepton masses.
This is done by solving the renormalization group (RG) equations for the Wilson coefficients
and matching the boundary conditions as one crosses each mass scale.
The results here can be used to relate CP violation quantities measured at high energies at colliders
to those at lower energies and vice versa. As an example some  leptonic processes cases are
studied in \cite{CN1}.

In  this paper we are more interested in processes below the weak scale.
In the next section we give a discussion of the evolution of the Wilson coefficients
from $\Lambda$ to  the weak scale and calculated the anomalous dimension of operators.
 Below the weak scale we integrate out the t-quark  and the heavy SM gauge bosons.
The RG running then  will involve only $U(1)_{em}$ and $SU(3)_c$  and will be
discussed in detail. We choose the low energy boundary condition at $m_p=1$ GeV.
 In section III  we use the effective Hamiltonian and fold in the appropriate form factors
to estimate the strength of signatures for  CPV experiments in $K$ meson  and pion
as well as semileptonic decays, and  the electric dipole moments of lepton and neutron.
They give stringent limits on the scalar and tensor coefficients.
As an example of the synergy between precision low energy measurements and high energy collider experiments
we calculated the CPV triple momentum-spin-spin correlation in $e^+ e^{-} \ra t+\bar{t}$ at the
linear collider.
 Our results are  summarized in the concluding Sec.IV.

\begin{figure}[h]
      \begin{center}
    \epsfig{file=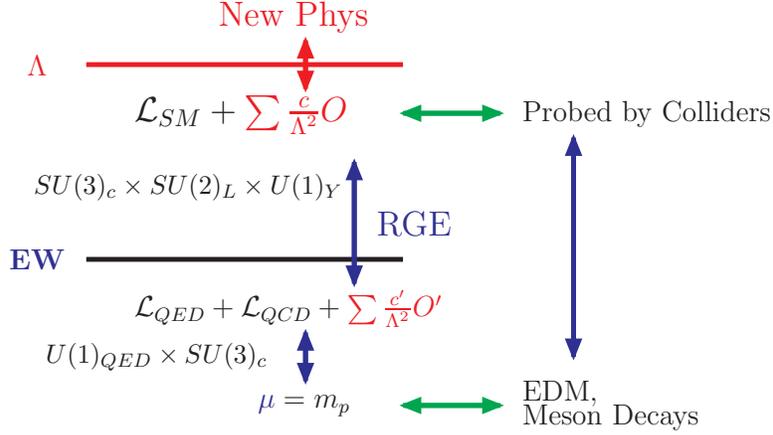, width=0.6\textwidth}
    \end{center}
  \caption{The connection between high and low energy interactions in effective
  field theory. The symmetry at each mass scale is given. All notations are given
  in the text. }
\label{scheme}
\end{figure}
A schematic representation of this approach is given in Fig.(\ref{scheme}). Although we have made the very
conservative assumption of no new states below $\Lambda$ the effective operative approach can be easily
generalized. For example if there were sterile neutrinos below $\Lambda$ the list of additional lepton operators
is given in \cite{CN1}. If the minimal supersymmetric standard model turns out to be the correct description of
physics just beyond the standard model then $\Lambda$ can be identified with the supersymmetry breaking
scale. Moreover, the operator list will have to be extended to include supersymmetric
state or states and
care has to be taken to respect the $R$-parity.

\section{ Effective Hamiltonian for Leptonic and Semileptonic Meson Decays}

We begin by calculating the quark level effective Hamiltonian for $d_i\ra u_j l \nu$.
This  given by the low energy SM with the W boson, Higgs boson, and heavy quarks integrated out
plus the relevant terms in Eq.(\ref{LCC}) with the Wilson coefficients taken at some energy $\mu$
below the weak scale. The term
$C_V^{\prime}$ is expected to make a undetectable correction to the Fermi coupling constant $G_F$ since
we are taking $\Lambda$ to be TeV or higher.  Thus, to leading order in weak
interactions and the new physics scale $\Lambda$ the effective Hamiltonian is given by
\beqa
\label{LSL}
{\cal{H}}_{\mathrm {eff}}^{s.l.}(\mu)&=&4\Gf  A(\mu)(V_{ij}P_{lk} \bar{u}_i\gamma^{\mu}\LH d_j)
(\bar{l}_l\gamma_{\mu}\LH \nu_k)
 \nonr \\
&& + \frac{C_{S1}^{ij,kl}(\mu)}{\Lambda^2}(\bar{u}_i\RH d_j)(\bar{l}_k\LH \nu_l)
-\frac{C_{S2}^{ij,kl}(\mu)}{\Lambda^2}(\bar{d}_i\RH u_j)(\bar{\nu}_k\RH l_l) \nonr \\
&& -\frac{C_T^{ij,kl}(\mu)}{\Lambda^2}(\bar{d}_i\sigma^{\mu\nu}\RH u_j)(\bar{\nu}_k\sigma_{\mu\nu}\RH l_l)  \nonr \\
&& +h.c.
\eeqa
where $V_{ij}$ is an element of the CKM matrix and $P_{lk}$ is an element of the neutrino mixing matrix.
The Wilson coefficient $A$ is  unity in the free quark limit and  are modified by SM gauge interactions as
ones goes from $\Lambda$ to the electroweak scale and then further down. The QED and QCD
running of $A(\mu)$ is well known whereas similar behaviors of scalar and tensor coefficients,
to the best of our knowledge,
have not been presented before. Hence we shall briefly discuss  below how
one can obtain the leading logarithmic (LL) contributions by renormalization group methods.
This also serves to establish our notations.

The operators $O_{S,V,T,D}$ which we shall generically denote by $O^6_A$ are bare operators.
 Upon renormalization they  will mix via the equation
\beq
O^{6}_A=\sum _{B=1}^{10} Z^{-1}_{AB}(Z_L)^{n/2}(Z_e)^{m/2}(Z_Q)^{r/2}(Z_q)^{s/2}\, O^{\prime 6}_B
\eeq
where  $Z_L,Z_e, Z_Q$ and $Z_q\; (q=u,d)$ are the wave function renormalization constants for the
various fermion fields.
$n,m,r, s $ are the number of such fields in each of the $O_A$  operators.
Thus, $n,m,r,s = 0,1,\, {\mathrm {or}}\; 2$.
The sum  here runs over the 10 terms  of $O_S, O_V$ and $O_T$ and the prime
denotes renormalized quantity.

 The renormalized operator $O^{\prime 6}_B$ will depend on the t'Hooft renormalization
 scale $\mu$ whereas the bare operators do not. Correspondingly the $\mu$-dependence of
 the Wilson coefficients will be such as to render the renormalized effective Lagrangian
 ${\cal L}^{\prime 6}$ independent of $\mu$. This leads to the
renormalization group equation (RGE) for the coefficients in Eqs.(\ref{L6mass}):
\beq
\label{RGC}
\mu\frac{d}{d\mu}C_A+\sum_B \gamma_{AB}C_B =0
\eeq
where $\gamma_{AB}$ is the anomalous dimension matrix which is non-diagonal. The values of $C_A$ at any scale
 $m$ are  obtained by solving the above equation plus boundary conditions which are the values given at another scale say the weak scale $M_W$. At present there are  only limits on a few of these coefficients from LEPII measurements.  We now return to discuss  $\gamma_{AB}$ calculation.

\begin{figure}[h]
  \centering
    \includegraphics[width=0.8\textwidth]{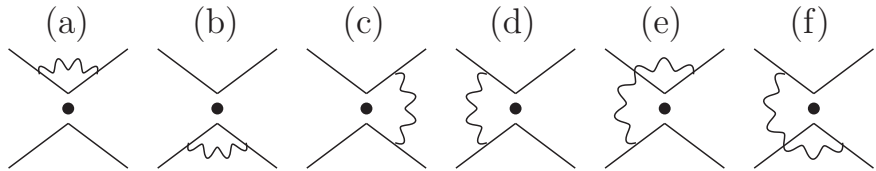}
  \caption{ 1-loop corrections to the 4-Fermi effective operators by SM gauge bosons represented by the wavy lines.
   }
  \label{fig:opmix2}
\end{figure}

Below the weak scale the renormalization of the operators are given by photon and gluon exchanges. We will
ignore all Yukawa couplings except that of the t-quark. The calculation becomes the evaluation of
 the high energy parts of the depicted diagrams in Fig.
 \ref{fig:opmix2}.
The wave function renormalization graphs also must be calculated but not shown. A  lengthy calculation gives
the following result for the 4-Fermi operators anomalous dimension matrix:
\beq
\gamma_{4F}={1\over 4\pi }
\left(\begin{array}{cccccccc}
G_1&&&&&&&\\
    & 2\alpha_1&0&0&0&0&0&\\
    &0&-2\alpha_1&0&0&0&0&\\
   &0&0& 4\alpha_1&0&0&0&\\
    &0&0&0& 4\alpha_1&0&0&\\
    &0&0&0&0&-8\alpha_1&0&\\
    &0&0&0&0&0&-\frac83\alpha_1\!-8\!\alpha_3&\\
    &&&&&&&G_2
\end{array} \right)\nonr
\eeq
in the basis of $\{c_{V1..V7},c_{S1},c_{S2},c_T\}$.
And $G_{1,2}$ are the two by two matrices:
\beqa
G_1&=& \left(\begin{array}{cc}-\alpha_1-3\alpha_2 &8\alpha_2\\
 6\alpha_2&-\alpha_1-7\alpha_2
\end{array} \right )\,,\\
G_2&=& \left(\begin{array}{cc}-\frac{11}{3}\alpha_1-8\alpha_3 &30\alpha_1-6\alpha_2\\
 \frac{5}{8}\alpha_1+\frac{3}{8}\alpha_2&\frac29\alpha_1+\frac83\alpha_3
\end{array} \right )\,.
\eeqa
After diagonalization and putting this into Eq.(\ref{RGC}) and solving it gives
the running of the scalar and tensor Wilson terms between $\Lambda$ and $M_W$.


To get a qualitative feeling of the RG running above the electroweak scale we first take the  running of
$SU(2)$ and $U(1)$  to be  much
smaller than from $SU(3)$ when the cutoff is below $10$ TeV. This is in accordance to usual expectation.
Considering  the
$SU(3)$ running alone leads to following simple solution.
Suppressing indices we define the enhancing factor
\beq
{\cal G}(M_W, \Lambda) \equiv {C(\Lambda) \over C(M_W)}\,.
\eeq
Then we have
\beq
{\cal G}_{V1..V7}(M_W, \Lambda) \sim   1.0 \pm {\cal O}(0.01)
\eeq
since only $U(1)$ and $SU(2)$ running play a role here. On the other hand QCD running affects the other
coefficients and we have
\beqa
&&{\cal G}_{S1}(M_W, \Lambda) \sim {\cal G}_{S2}(M_W, \Lambda) \sim
\left( {\alpha_3(M_W) \over \alpha_3(M_t)}\right)^{-\frac{12}{23}}
\left( {\alpha_3(M_t) \over \alpha_3(\Lambda)}\right)^{-\frac{4}{11}}\,,\\
&& {\cal G}_T (M_W, \Lambda) \sim  \left( {\alpha_3(M_W) \over \alpha_3(M_t)}\right)^{\frac{4}{23}}
\left( {\alpha_3(M_t) \over \alpha_3(\Lambda)}\right)^{\frac{4}{33}}\,.
\eeqa
At $\Lambda= 1$ TeV, we find that ${\cal G}_{S1,2}=0.865$ and ${\cal
G}_T=1.049$. We can also give approximate expressions for the above quantities :
\beq
\label{eq:RGEST2Lam}
{\cal G}_{S1,2}(M_W, \Lambda)\sim 0.9456 \left({m_t \over \Lambda
}\right)^{0.036}\,,\;
{\cal G}_T(M_W, \Lambda)\sim 1.0188 \left({\Lambda \over m_t }\right)^{0.012}
\eeq
where the number factors in front of the parentheses are the result
of RG running from $M_W$ to $m_t=174$ GeV.

Below the EWSB scale the operating gauge symmetry is $U(1)_{em}\times SU(3)_c$.
It is well known that there are no large LL contribution from QCD for semileptonic decays in the SM.
The perturbative QCD corrections have been calculated and is found to be small \cite{Stein}; henceforth,
we shall neglect it in $A(\mu)$.
However, QED gives rise to LL corrections in $A(\mu)$ \cite{MS} and the beta function is
obtained from a subset of the diagrams listed in Fig.\ref{fig:opmix2}.
Explicitly we have
\beq
A(m_i,m_W)=
\left[\frac{\alpha(m_c)}{\alpha_(m_i)}\right]^{\frac{3}{8}}
\left[\frac{\alpha(m_\tau)}{\alpha_(m_c)}\right]^{\frac{9}{32}}
\left[\frac{\alpha(m_b)}{\alpha_(m_\tau)}\right]^{\frac{9}{38}}
\left[\frac{\alpha(M_W)}{\alpha_(m_b)}\right]^{\frac{9}{40}}
\eeq
where $i=d,s$.
 In the above we  have integrated  out simultaneously  the t-quark and the gauge bosons and then the light fermions in succession. $m_i$ is either the strange quark mass or the proton mass and we have noramlized $A(m_i,m_i)$ to unity. Using the values
\beqa
\alpha^{-1}(M_W)&=& 128.0\,, \nonr \\
\alpha^{-1}(m_b)&=& 132.14\,, \nonr \\
\alpha^{-1}(m_{\tau})&=& 133.33\,, \nonr \\
\alpha^{-1}(m_c)&=& 133.69\,, \nonr\\
\alpha^{-1}(m_p)&=& 133.91\,,
\eeqa
where we choose $m_b=4.3$ GeV, $m_c=1.3$ GeV, and  $m_p\sim 1$ GeV. We find
\beq
A(m_p,M_W)= 1.0107\,,
\eeq
which is an important correction for precision measurements.

For $C_{S1}$, we find
\beq
\gamma_{S1}= {1\over 4\pi}\left( \frac43 \alpha -8 \alpha_3
\right)\,,
\eeq
 and have the solution:
\beqa
C_{S1}(m_p)=
\left[\frac{\alpha(m_c)}{\alpha(m_p)}\right]^{-\frac{1}{8}}
\left[\frac{\alpha(m_\tau)}{\alpha(m_c)}\right]^{-\frac{3}{32}}
\left[\frac{\alpha(m_b)}{\alpha(m_\tau)}\right]^{-\frac{3}{38}}
\left[\frac{\alpha(M_W)}{\alpha(m_b)}\right]^{-\frac{3}{40}}\nonr\\
\times
\left[\frac{\alpha_3(m_c)}{\alpha_3(m_p)}\right]^{-\frac{4}{9}}
\left[\frac{\alpha_3(m_b)}{\alpha_3(m_c)}\right]^{-\frac{12}{25}}
\left[\frac{\alpha_3(M_W)}{\alpha_3(m_b)}\right]^{-\frac{12}{23}}
C_{S1}(M_W)\,.
\eeqa
Using the boundary condition $\alpha_s(M_W)=0.120$ and the standard 1-loop QCD beta function,
we get
\beqa
\alpha_3(m_b) &=& 0.211\,,\nonr\\
\alpha_3(m_c) &=& 0.316\,,\nonr\\
\alpha_3(m_p) &=& 0.359\,,
\eeqa
and
\beq
C_{S1}(m_p)= (0.9965)\times (1.7232) \times C_{S1}(M_W)\,,
\label{eq:CS1RGE}
\eeq
where the first and second brackets are the QED and QCD RG running
effect respectively. We see that the RG effects enhance the coefficient
almost by factor 2.

The running of  $C_{S2}$ and $C_{T}$ are more complicated since they are coupled:
\beqa
\label{CS2T}
\gamma_{ST} &=& {\alpha \over 4\pi} \left( \begin{array}{cc}4/3  & 8 \\ 1/6& -40/9 \end{array}\right)
+ {\alpha_3 \over 4\pi} \left( \begin{array}{cc}-8 & 0\\0& 8/3 \end{array}\right)
\eeqa
in the basis of  $ \{ C_{S2}, C_T\}^T$.

An analytic solution for Eq.(\ref{CS2T}) is unavailable. For a solution numerical methods are required.
However, since  the QED effect is expected to be much smaller than
QCD; for simplicity, we can just keep the QCD part to have a rough idea how the
Wilson coefficient evolves. Using the same parameter setting as in Eq.(\ref{eq:CS1RGE}), we get
\beq
C_{S2}(m_p) \sim 1.72\, C_{S2}(M_W)\,,\; C_{T}(m_p) \sim 0.83\,  C_{T}(M_W)\,,
\label{eq:RGECS2TW}
\eeq
which should be sufficient for order of magnitude estimates of new physics effects. It is interesting to note
that the scalar coefficients increases as the energy decreases whereas the tensor one does the opposite.

\section{Phenomenology}
\subsection{Leptonic Decays of K and $\pi$ Mesons}

We can now apply the above formalism to leptonic decays of pseudoscalar mesons. In particular
the branching
ratios $R_{\pi}=B(\pi\ra e \nu/\pi\ra \mu \nu)$ and $R_K=B(K\ra e \nu/K\ra \mu \nu)$ which are precisely
measured. Theoretically they have been precisely calculated in the SM \cite{SMRPI} with the uncertainties
below the experimental limits; thus making these decays  very sensitive tests of new
physics and in particular scalar interactions.

To calculate the amplitude a pseudoscalar meson M ($\pi^+, K^+ \mathrm{or} \, B^+)$ decays into a lepton pair
we need to introduce the following form factors \cite{PDG}
\beqa
\langle 0|\bar{u}_i\gamma^{\mu}\gamma_5 d_j|M(p)\rangle &=&i f_M p^{\mu}\,, \nonr \\
\langle 0|\bar{u}_i\gamma_5 d_j|M(p)\rangle &=& i f_M \left( M^2/(m_i+m_j)\right)\,,
\label{form}
\eeqa
where $M$ is the meson mass. The pseudoscalar form factors take the values $f_{\pi}= 130.7(4)$ MeV and
$f_K=160(2)$ MeV although we do not
need these values. On the other hand $f_B$ is not known but expected to be about 200 MeV.

Since the pseudoscalar meson carries only one momentum, $p$.
It can't have an antisymmetric form factor due to  Lorentz
invariance,
\beqa
\langle 0|\bar{u}_i \sigma^{\mu \nu}d_j| M(p)  \rangle &=&0 \,,\nonr \\
\langle 0|\bar{u}_i \sigma^{\mu \nu}\gamma_5 d_j| M(p) \rangle &=& 0\,.
\eeqa
However, one can have  nonzero form factors for tensor interaction if there
is an extra particle in the final state such in  the $\pi^- \ra e \nu
\gamma$ case \cite{Herczeg:1994ur} to which we shall return at the end of the section.

Substituting Eq.(\ref{form}) into Eq.(\ref{LSL})
the amplitude for the decay $d_j\ra u_i l \nu_k$ for a specific charged lepton $l$ but
sum over all three active neutrinos is then given by
\beqa
i{\cal M}&=&\left. \frac{2G_F}{\sqrt 2} f_M\right[  V_{ij} A(m_i)
\left(P_{lk} \bar{l}\, p^\mu \gamma_{\mu}\LH\, \nu_k \right)  \nonr\\
&& - \left. \frac{\sqrt 2 M^2} {4 G_F\Lambda^2(m_i+m_j)}
\left(C^{ij,lk}_{S1}(m_i)-C^{ij,kl *}_{S2}(m_i)\right)\left( \bar{l} \LH \nu_k \right) \right]\,,
\eeqa
and notice that only a sum over $k$ is taken. From this  we obtain
\beq
R_{\pi}= R_{\pi}^{SM} \times
\left( 1  +{ K^e_\pi - K_\pi^\mu \over  G_F\Lambda^2 }   \right)
\label{pilnu}
\eeq
where
\beq
R_{\pi}^{SM}= { m_e^2 (M_{\pi}^2-m_e^2)^2 \over m_\mu^2 (M_{\pi}^2 -m_{\mu}^2)^2 }
\left(1 -16.1\, \frac{\alpha}{\pi}\right) =1.2354(2)\times 10^{-4}
\eeq
is the SM prediction \cite{SMRPI}, and
\beq
\label{Kpi}
K^l_\pi= {1 \over \sqrt 2 A(m_i,M_W) V_{ud}  }
 { M_{\pi}^2 \over m_l(m_u+m_d)}
 \sum_k{\frak{Re}}\left[P^*_{kl}\left( C_{S2}^{11,k l*}- C_{S1}^{11, l k}\right)\right]\,.  \nonr
\eeq
Because we have factored out the SM helicity suppression factor in Eq.(\ref{pilnu}) the correction factor
given by Eq.(\ref{Kpi}) has a $1/m_l$. This is a reflection of the helicity flipping nature of
scalar interactions.

Compared with the data $R_{\pi}^{exp}=1.230(4)\times
10^{-4}$\cite{PDG} which is consistent with the SM
we have the limit:
\beq
\label{pil2}
\left| { K^e_\pi - K_\pi^\mu \over  G_F\Lambda^2 } \right| \leq
0.007\; ( 68\%\, \mbox{C.L.} )
\eeq

Besides the overall scale factor $\Lambda$ for the new physics, Eq.(\ref{pil2}) gives valuable information
of these scalar coefficients. The first observation is that the phases of $C_{S1}^{ij,kl}$ and
$C_{S2}^{ij,kl}$ must not accidentally cancel for  it to be useful. Next we  define
\beq
C_S^{\pi,kl}\equiv C_{S1}^{11,kl*}-C_{S2}^{11,kl} \;\;\; (l= e , \mu)
\eeq
 and consider the following general possibilities which cover most of the relevant models for this problem:
\benu
\item
There is no hierarchy (NH) between the electron and muon  modes.

An an  example we take
\beq
\label{nohier}
C_S^{\pi,ke}\sim C_S^{\pi,k\mu}\,.
\eeq
Then $K_{\pi}^e$ dominates in Eq.(\ref{pil2}) and taking $M_\pi/(m_u+m_d)=15$, we have:
\beq
\label{cslimit}
\left| \sum_k {\frak {Re}}\left[P^*_{ke}\left(C_{S2}^{11,k e*}- C_{S1}^{11, e k}\right)\right] \right|_{
\mathrm {NH}}
\leq 2.8 \times 10^{-5}  \left( {\Lambda \over  \mbox{TeV}}\right)^2\; ( 68\%\, \mbox{C.L.} )
\eeq
\item
An exact hierarchy exists between the electron and muon modes.

As an illustration we take the
coefficients to be proportional to the electron and muon Yukawa couplings, $y_e\sim 3\times 10^{-6}$ and
$y_{\mu}\sim 6\times 10^{-4}$ respectively.
Such is the case for two Higgs doublet model (2HMD). Explicitly we have
\beqa
\label{exhier}
C_S^{\pi,k e}&=&y_e C\,, \nonr \\
C_S^{\pi,k \mu}&=& y_{\mu}C\,,
\eeqa
where $C$ is an undetermined constant which for simplicity we take to have no dependence on the neutrino indices.
In this case the constraint from Eq.(\ref{pil2}) becomes
\beq
\left| C\sum_k {\frak{Re}} \left[ P_{ke} -P_{k\mu} \right] \right|
\leq 9.3 \left( {\Lambda \over
\mbox{TeV}}\right)^2\,.
\eeq
Clearly the constraint from lepton universality tests is not as stringent as the previous case;
nevertheless is still an interesting one.
Furthermore,
it also demonstrates the importance of taking into account neutrino mixings.
\eenu

The above analysis can be repeated  for K decays. We have
\beq
\label{klnu}
R_{K}= R^{SM}_{K}\times
\left( 1  +{ K^e_K - K_K^\mu \over  G_F\Lambda^2 }   \right)
\eeq
where $R_{K}^{SM}$ is the SM prediction and
\beq
\label{Klnu}
K^l_K= {1 \over  \sqrt 2 A(m_i,M_W) V_{us}  }
{ M_{K}^2 \over m_l(m_u+m_s)}
 \sum_k {\frak{Re}}\left[P^*_{k l}\left(C_{S2}^{21,kl *}- C_{S1}^{12, l k}\right)\right]. \nonr
\eeq
The most recent data from NA48/2 \cite{NA48} is
\beq
\label{NA48}
R_K^{\mathrm{exp}}=(2.416\pm 0.043_{stat}\pm 0.024_{syst})\times 10^{-5}
\eeq
which improves upon the PDG value \cite{PDG} :
$R_{K}^{exp}=2.44(11)\times 10^{-5}$.
 This is to be compared with the SM value of $R_K^{SM} = 2.472(1)\times 10^{-5}$ \cite{SMRPI}.
Using $M_K/(m_u+m_s)=3$ and $V_{us}=0.224$, we have the limit:

\beq
\label{kl2limit}
\left| \sum_k {\frak{Re}}\left[P^*_{k e}\left(C_{S2}^{21,k e *}- C_{S1}^{12, e k}\right)\right] \right|_{\mathrm{NH}}
\leq 6.4\times 10^{-5}  \left( {\Lambda \over
\mbox{TeV}}\right)^2\; ( 68\%\, \mbox{C.L.} )
\eeq
For the exact hierarchy case we get
\beq
\left| C^{\prime}\sum_k{\frak{Re}}\left[ P_{ke}-P_{k\mu}\right]\right|\leq 21.3 \left( {\Lambda \over  \mbox{TeV}}\right)^2
\eeq
where $C^{\prime}$ replaces $C$ of Eq.(\ref{exhier}).

As remarked earlier we have to use the radiative decays to get constraint
on the tensor coefficients. However,  as has been pointed out by \cite{Voloshin:1992sn},
due to QED correction the tensor operator will induce scalar
operator. And it's been widely believed  that
the limit on $C_T$ from $\pi^-\ra e\nu$ is two orders better than the
direct constraint from $\pi^-\ra e\nu \gamma$.

\begin{figure}[h]
      \begin{center}
    \epsfig{file=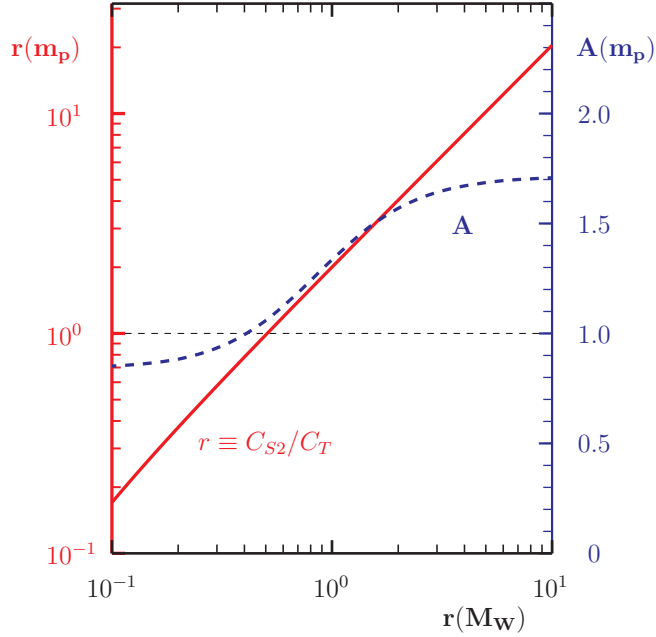, width=0.6\textwidth}
    \end{center}
  \caption{The full RG running for the coupled scalar and tensor Wilson coefficients
  from $M_W$ to $m_p\sim 1$ GeV. The x-axis is the ratio
  $r=C_{S2}/C_{T}$ at $M_W$, and the left-handed y-axis is the ratio $r$ at $m_p$.
  The dash line (blue) is  the amplitude at low energy, we set $A\equiv \sqrt{|C_{S2}|^2+|C_T|^2}=1$ at $M_W$,
  refer to the axis at the right-handed side.   }
\label{fig:RGS2T}
\end{figure}

Since we have a RG improved  result given in Eq.(\ref{CS2T}),
it's interesting to see how this affects the above conclusion. To that end
we carried out a full numerical analysis for the coupled
equations Eq.(\ref{CS2T}) and the result is displayed in
Fig.\ref{fig:RGS2T} where  the figure caption explains our notations.

Assuming there is no hierarchy between the Wilson coefficients $C_{S2}$ and $C_T$ at EW scale,
for instance the two are within one order of magnitude.
Basically, we find the RG evolution  can be simply summarized  by Eq.(\ref{eq:RGECS2TW}).

Let us  consider the extreme examples that $(C_{S2}, C_T)=(0,1)$ and $(1,0)$ at
EW scale. The first corresponds to only tensor interactions are induced by new physics
whereas the second has only scalar being generated. Our numerical solution gives $(-0.03, 0.84)$
and $(1.717, -5\times 10^{-4})$ at $m_p$ respectively.
In other words, if the new physics beyond SM gives only  tensor interaction at the weak scale; then
 at  meson mass scale the RG evolution gives
\beq
\label{csoverct}
{ C_{S2}(m_p) \over C_T(m_p)} = -0.036\,.
\eeq
This  is to be compared with the estimation given in \cite{Voloshin:1992sn}
\beq
{ C_{S2}(m_p)\over C_T(m_p)} = -\frac{\alpha}{\pi} \ln\frac{M_W^2}{m_p^2}
= -0.020\,.
\eeq
We see the full RG running gives about $80\%$ correction. Using Eq.(\ref{cslimit}) and Eq.(\ref{csoverct})
we get the upper limit
\beq
\left|{\frak{Re}}  C_T(m_p)\right|< 7.7 \times 10^{-4}  \left( {\Lambda \over
\mbox{TeV}}\right)^2
\eeq
which is a factor of $\sim 250$ better than direct measurement of $\pi\ra e\nu \gamma$.

\subsection{Polarization in $K^+\ra \pi^0\mu^+ \nu$}

It is well known that the transverse muon polarization in semileptonic pseudoscalar mesons decays are sensitive
 CP violation tests
of  effective scalar interactions \cite{BG, WN}. We revisit these studies by adding the RG
running of the effective operators to the previous studies which were conducted mostly in the context of
specific models \cite{FV,GK}.

The effective Hamiltonian at the quark level is the same as in Eq.(\ref{LSL}).
The form factors for $K^{+}(p_K)\ra \pi^0(p_{\pi})\mu^+(p_{\mu}) \nu(p_{\nu})$ decay that we need are
\beqa
\label{kl3f}
&\langle \pi^0|\bar{s}\gamma^{\mu}u|K^+ \rangle & =  f_{+}p_{+}^{\mu}+f_{-}p_{-}^{\mu}\,, \nonr \\
&\langle \pi^0|\bar{s} u|K^+ \rangle & \simeq f_{+}\frac{M_K^2-M_{\pi}^2}{m_s}\,, \nonr \\
&\langle \pi^0|\bar{s}\sigma^{\mu\nu}u|K^+ \rangle & \simeq
i\frac{f_T}{M_K}\left(p_{K}^{\mu}p_{\pi}^{\nu}- p_{K}^{\nu}p_{\pi}^{\mu}\right)\,,
\eeqa
where $p_{\pm}=(p_K\pm p_{\pi})$ and  we have  neglected  the u-quark mass.
Again, the form factors of $(\bar{s}\gamma^{\mu}\gamma_5u)$, $(\bar{s}\gamma_5 u)$, and
$(\bar{s}\sigma^{\mu\nu}\gamma_5u)$ vanish due to parity.
 The form factors are functions of $p_{-}^2$ and is normalized to $f_{+}(0)=1$.
 Furthermore, one usually defines  $\xi\equiv \frac{f_{-}}{f_{+}}$.
 In the second of Eq.(\ref{kl3f}) we have omitted a term proportional
to $\xi$ by taking advantage of its small value; i.e. $|\xi|=0.124$ \cite{PDG}.
In the third of Eq.(\ref{kl3f}), we have dropped a term
$\delta f_T\times (\epsilon^{\alpha\beta\mu\nu}p_{K \alpha} p_{\pi \beta})$
which can be taken into account  by redefining  $f_T\ra (f_T -\delta f_T)$ when
contract with the lepton current part.
 The decay amplitude is
\beqa
\label{Kl3m}
-{\cal M} &=&\left. \sqrt{2}G_Ff_{+}\right[ V^*_{us}P_{2k}A(m_s)(p_{+}^{\alpha}+\xi p_{-}^{\alpha})\bar{\nu}\gamma_{\alpha}
\LH\mu \nonr\\
&&+\frac{(M_K^2-M_{\pi}^2)}{2\sqrt2 G_F\Lambda^2 m_s}\left( C_{S1}^{*12,2k}(m_s)-C_{S2}^{21,k2}(m_s)\right)
 \bar{\nu}_k \RH\mu \nonr \\
&&-\left. i\frac{f_T}{f_{+}}\frac{C_T^{21,k2}(m_s)}{\sqrt 2 G_F\Lambda^2M_K}p_K^{\alpha}p_{\pi}^{\beta}
\bar{\nu}_k\sigma_{\alpha\beta}\RH\mu\right]\,.
\eeqa
The above can be easily compared with the standard notation
\cite{PDG}. We further define two variables
\beqa
\xi_S &\equiv& {1 \over 2\sqrt{2} V_{us} A(m_s)}{M_K^2-m_\pi^2 \over M_K m_s }{ C_S^K\over \Lambda^2 G_F}\,,\\
\xi_T &\equiv& -{f_T/f_+  \over \sqrt{2} V_{us} A(m_s)} {C_T^K \over  \Lambda^2 G_F }\,,
\eeqa
where
\beq
C_S^K= \sum_k P^*_{k \mu} \left(
C_{S1}^{*12,2k}(m_s)-C_{S2}^{21,k2}(m_s)\right)\,,
\eeq
and
\beq
C_T^K=\sum_k P^*_{k \mu} C_T^{21,k2}(m_s)\,.
\eeq
The CP violating transverse muon polarization is found to be:
\beq
P_\bot \sim \left[\frak{Im} \xi_s + {p_K\cdot(p_\nu-p_\mu)+m_\mu^2/2 \over M_K^2} \frak{Im} \xi_T \right]
 {\vec{p}_\mu \times \vec{p}_\nu \over \Phi}
\eeq
where $\vec{p}$'s are  leptons' 3-vector momentum in the
kaon rest frame and we have ignored the correction of higher power in $\xi_T$
and
\beqa
\Phi &=&(2E_\mu M_K-m_\mu^2)\left(1-{E_\pi+E_\mu \over
M_K}\right)\nonr\\
&-&\frac12(M_K^2+m_\pi^2-m_\mu^2-2E_\pi
M_K)\left(1-\frac{m_\mu^2}{4M_K^2}\right)\,.
\eeqa
Plugging in the numbers, we have
\beq
P_\bot \sim \left[ 0.38 \frak{Im} C_S^K - 0.27{p_K\cdot(p_\nu-p_\mu)+m_\mu^2/2 \over M_K^2( f_+/f_T)}
 \frak{Im} C_T^K \right]\left( {\mbox{TeV}\over \Lambda}\right)^2
 { \vec{p}_\mu \times \vec{p}_\nu \over \Phi}\,.
\eeq
It is interesting to note the imaginary part of  both tensor and scalar
coefficients are sensitive to where measurements are made in
the  Dalitz plot distribution \cite{Geng_Lee}. $C_T$ has maximum sensitivity when $E_{\mu}-E_{\nu}$ is largest
in the kaon rest frame .
From the current limit $|P_T|<0.0050$ \cite{KEKE246} and assuming that $f_T \simeq f_+$, we have
\beq
\label{Imclimit}
\left| \frak{Im} C_S^K \right|\, \mbox{and}\, \left| \frak{Im} C_T^K  \right|
\leq 2\times 10^{-3} \left( { \Lambda \over
\mbox{TeV}}\right)^2\,.
\eeq
Notice here that  the results from $K_{l2}$ decays
limits  the real part of the scalar coefficients.

\subsection{Constraints from Electron and Neutron  EDM's}
It is well known that the EDM of the electron is a very sensitive test of new physics phases. This is
due to the fact that the SM gives an undetectably small value $|d_e|\leq 10^{-38}$ e-cm. In addition
there are several ongoing experimental efforts to reach the level of $10^{-30}$. If discovered it will
be a clear sign of new physics. Similarly the neutron EDM, $d_n$,  is also very sensitive to new physics phases
and are usually connected with electron EDM in model dependent ways. Hence, they are very complementary
tests of new physics. Although the SM prediction for $d_n$ is not as robust as $d_e$ but is generally taken to
be $< 10^{-31}$ e-cm. This is still very much below the experimental bound of \cite{PDG}
\beq
\label{dnbound}
|d_n^{\mathrm{exp}}| < 6.3 \times 10^{-26} \; \mbox {e-cm}\,.
\eeq

Among the terms in the effective Lagrangian  new phases reside in the scalar and tensor
Wilson coefficients in ${\lag}_6^{\mathrm {NC}}$ and ${\lag}_6^{\mathrm {CC}}$. Some  NC terms will
contribute to $d_l$ at the 1-loop level and the CC term will induce it at the 2-loop level. This is explored
in detail below.

\subsubsection{EDM at One Loop}
\begin{figure}[h]
      \begin{center}
    \epsfig{file=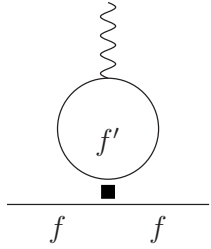, width=0.2\textwidth}
    \end{center}
  \caption{1-loop contribution to the EDM of a charged fermion $f$ from semileptonic 4-Fermi operators denoted  by
the box and $f$ represents $l$ or a u-quark and $f^{\prime}$ a u-quark or $l$ respectively.}
\label{fig:1lpedmt}
\end{figure}
The generic one loop diagram for $d_f$ is given in Fig.(\ref{fig:1lpedmt}). It has been shown
\cite{Degrassi:1998es} that the RG running of $d_l$ below the EW scale is not significant and
we shall ignore it. For the quark EDM, QCD correction is the biggest contribution.
We found its QCD anomalous dimension is same as  the
tensor operator which we have already found its solution in Eq.(\ref{eq:RGECS2TW}), i.e.
\beq
d_u (m_p) \sim 0.83 d_u(M_W)\,.
\eeq
Hence, for our purpose we can neglect the RG running in both cases.

It has been shown \cite{CN1}
that scalar operators will not contribute at this level. The vector terms in Eq.(\ref{LNC}) have real coefficients
and hence will play no role. This leaves the terms $C_T^{ii,ee}$ and only the u-type quarks come into play.
A straightforward calculation leads to
\beq
d_e= \frac{e N_c }{3\pi^2\Lambda^2}
\sum_{i=u,c, t} m_i \frak{Im} C_T^{ii,ee}\ln\frac{\Lambda^2}{m_i^2}
\eeq
at the EW scale which is where  we set the boundary condition for the RGE of EDM
operator and $N_c$ is the number of colors.  For simplicity, we ignore the top threshold effects
and the RG evolution between $m_t$ and $M_W$.
Similarly, we have
\beq
d_u= - \frac{e }{2\pi^2\Lambda^2}
\sum_{l=e,\mu, \tau} m_l \frak{Im}
C_T^{11,ll}\ln\frac{\Lambda^2}{m_l^2}\,.
\eeq
Notice that there is no $d_d$ at this level due to the $SU(2)_L$ symmetry.

Clearly the most important term is due to the t-quark. This has the unusual behavior of depending on the internal
fermion mass and independent of $m_e$. The chirality changing nature of tensor interactions is responsible.
Thus, one expects the same formula to hold  for $d_{\mu}$ and $d_{\tau}$ with obvious
change to the lepton indices.
>From the experimental limit  of $|d_e|\leq 1.7 \times 10^{-27}$ e-cm \cite{eedm}
 we deduce that
\beq
\label{eq:boundT}
\left| \frak{Im} C_T^{33,ee} \right| \leq 1.4\times 10^{-9} \left( { \Lambda \over \mbox{TeV}}\right)^2
\eeq
which is much more stringent than one obtains from meson decays; see Eq.(\ref{Imclimit}),  albeit for different
flavor indices.

Assuming that the leading contribution to $d_n$ comes from $d_u$
and using the quark model relation $d_n\sim (4 d_d-d_u)/3$ for estimation, we obtain from Eq.(\ref{dnbound})
a less stringent bound
\beq
\left| \frak{Im} C_T^{11,\tau\tau} \right| \leq 8.4\times 10^{-6} \left( { \Lambda \over
\mbox{TeV}}\right)^2\,.
\eeq

\subsubsection{EDM at Two Loops}
\begin{figure}[h]
      \begin{center}
    \epsfig{file=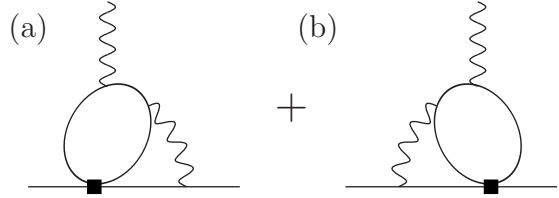, width=0.5\textwidth}
    \end{center}
  \caption{ 2-loop contribution to the EDM of a charged fermion $f$ from semileptonic 4-Fermi operators denoted  by
the box. The internal gauge boson line is either a photon or W-boson.  }
\label{fig:2lpedm}
\end{figure}

Unlike ${\lag}_6^{\mathrm {NC}}$ the complex coefficients in ${\lag}_6^{\mathrm {CC}}$ can only induce EDM
at the 2-loop level. Similarly the scalar coefficients of ${\lag}_6^{\mathrm{NC}}$ come into play now.
The generic diagrams for this are  displayed in Fig.(\ref{fig:2lpedm})
Now both the scalar terms $C_{S1,2}$ contribute. From the above discussions we can ignore the tensor terms
since they are constrained to be small (see also the next section).

An examination of ${\lag}_6^{\mathrm{NC}}$ and the structure of the Feynman diagrams shows that only
the flavor diagonal terms come into play. Since the lepton masses are small they can be neglected
in the calculation and we only need to keep the quark masses. Also the leading contribution
comes from internal photon exchange.
Then the NC contribution to the
 EDM $d_l$ of a charged lepton $l= e, \mu$ or $\tau$ is given by
\beq
d_l(\mathrm {NC}) = {e \alpha  \over 48 \pi^3\Lambda^2}
\left[ 4\sum_{i=u,c,t} m_i {\frak{Im}}C^{ii,ll}_{S2}{\cal F}_u\left( {\Lambda^2\over m_i^2}\right)
+ \sum_{j=b,s,t} m_j {\frak{Im}}C^{ii,ll}_{S1} {\cal F}_d\left( {\Lambda^2\over m_j^2}\right)
\right] \, ,
\label{EDMNC}
\eeq
 and the functions are
\beqa
{\cal F}_u(z)&=&  \int^z_0 dx \int^1_0 dy {1-y(1-y) \over 1+y(1-y)x}\,,\\
{\cal F}_d(z)&=&  \int^z_0 dx \int^1_0 dy {y(1-y) \over 1+y(1-y)x}\,.
\eeqa
These results extend our previous expression in \cite{CN1} where only
partial contribution was considered.
We have ${\cal F}_t=\{10.872, 59.552,150.780\}$ and ${\cal F}_b=\{8.596,13.200,17.808\}$
for  $\Lambda=\{1,10,100\}$ TeV. When $z\gg 1$, they can be
approximated by
\beqa
{\cal F}_t(z)&\simeq&  \left(2  - \ln z + \ln^2 z\right)\,,\nonr\\
{\cal F}_b(z)&\simeq&  -\left(2- \ln z\right)\,.
\eeqa

 There are two helicity flips in the above result. The first one
involves the scalar 4-Fermi interactions $C_{S1,2}$. The second one is the helicity flip from the
quarks in the loop due to the chiral nature of the weak interactions and  is explicitly displayed
 in Eq.(\ref{EDMNC}).
As expected, in the limit of massless fermions there is no EDM. It is also clear that only
the t and b-quarks are important.

For the contributions from ${\lag}_6^{\mathrm{CC}}$ the calculation is similar. Now the internal
gauge boson exchange comes from the  W boson. The result is
\beqa
\label{EDMCC}
d_l({\mathrm{CC}}) &\sim& {e \alpha  \over 16 \pi^3\sin^2 \theta_w \Lambda^2}
\left[ m_t \frak {Im}C_{S2}^{33,ll}{\cal G}_t\left({\Lambda^2\over M_W^2}\right)
+m_b \frak {Im}C_{S1}^{33,ll}{\cal G}_b\left({\Lambda^2\over M_W^2}\right)    \right]\,,\\
{\cal G}_t(z) &=& \frac18\int^z_0 d x \int^1_0
 dy {(2-3y)(2-y)x \over \left(m_t^2/M_W^2 +y x\right)(1+x)}\,,\\
{\cal G}_b(z) &=& -\frac18\int^z_0 d x \int^1_0
 dy {y(2-3y) x \over \left(m_t^2/M_W^2+y x\right)(1+x)}\,,
\eeqa
where we have kept only the  $t$ quark mass in the quark propagators. The analytic expression
for ${\cal G}$ is rather complicated and not illuminating.
From  numerical study we have ${\cal G}_t=\{1.708, 11.450, 31.670\}$
and ${\cal G}_b=\{-0.0824, -0.347, -0.6343\}$
for $\Lambda=\{1,10,100\}$ TeV.
For our purpose it is good enough to use the following approximation when $z\gg 1$:
\beqa
{\cal G}_t(z)&\simeq&  \frac14 \left( 13.7 -6.31 \ln z + \ln^2 z \right)\,,\nonr\\
{\cal G}_b(z)&\simeq&  \frac14\left( 1-\frac14 \ln z\right)\,.
\eeqa
Combining the above two contributions, we arrive at the estimation:
\beq
d_l \sim {e\alpha \over 16\pi^3 \Lambda^2}\left\{
m_t{\frak{Im}} C_{S2}^{33,ll}\left[\frac43 {\cal F}_t+{{\cal G}_t \over \sin^2\theta_w} \right]
+m_b{\frak{Im}} C_{S1}^{33,ll}\left[\frac13 {\cal F}_b+{{\cal G}_b \over \sin^2\theta_w} \right]
\right\}\,.
\eeq
The physics is now clear. For $C_{S1/S2}$, due to the helicity,
it can only pick up the bottom/top  quark mass in the loop.
Barring accidental  cancellation between  the two scalar coefficients terms
the upper limit of the imaginary parts can be deduced from electron EDM experiment:
\beqa
\label{eq:CSboundEDM}
\left|{\frak{Im}} C_{S2}^{33, ee} \right| <1.5 \times 10^{-6} \left({\Lambda\over
\mbox{TeV}}\right)^2\,,\nonr\\
\left|{\frak{Im}} C_{S1}^{33, ee} \right| <4.7 \times 10^{-4} \left({\Lambda\over
\mbox{TeV}}\right)^2\,.
\eeqa

The results from our one and 2-loop study show that EDM's are sensitive to the imaginary parts of the flavor
diagonal scalar coefficients. On the other hand the charged lepton polarization experiments  directly
probe the flavor off diagonal terms although at a less sensitive level.
\subsection{ Muon Anomalous Magnetic Moment, $a_{\mu}$ and Rare Decays.}
The exact  same 1-loop  calculation  leads to the following  effective Lagrangian:
\beqa
\triangle{\cal L} &\sim&  -{e m_t C_T^{33,ij} \over 2\pi^2\Lambda^2}
\ln\frac{\Lambda^2}{m_t^2} \left( \bar{e}_i \sigma^{\mu\nu}\RH e_j\right) F_{\mu\nu}
+h.c.\nonr\\
&& + {e m_\tau C_T^{ij,\tau\tau} \over 4 \pi^2\Lambda^2}
\ln\frac{\Lambda^2}{m_{\tau}^2} \left( \bar{u}_i \sigma^{\mu\nu}\RH u_j\right) F_{\mu\nu}
+h.c.
\eeqa
The flavor diagonal entries in the first term also contribute to $(g-2)_{\mu}$ and off diagonal terms lead
to the rare decay $(\mu\ra e+ \gamma)$.
Thus, the  branching ratio of the radiative decay of a charged lepton $L$  normalized to
 the decay width of $( L\ra l +\nu_L +\bar{\nu}_l)$ is given by \cite{CN2}:
\beq
B(L\ra l +\gamma) = 48  \frac{\alpha m_t^2}{\pi M_L^2}
 { \ln^2\frac{\Lambda^2}{m_t^2} \over (\Lambda^2 G_F)^2 }
\left( \left|C_T^{33,Ll}\right|^2 + \left|C_T^{33,lL}\right|^2
\right)\,.
\eeq
>From the limit $B(\mu\ra e \gamma)<1.2\times 10^{-11}$,  $B(\tau\ra \mu \gamma)< 6.5\times 10^{-6}$
and $B(\tau\ra e \gamma)<1.6\times 10^{-5}$ \cite{PDG}, we get
\beqa
\label{LFVT}
\left|C_T^{33,\mu e}\right|^2 + \left|C_T^{33,e\mu}\right|^2 <
4.0\times 10^{-16} \left( {\Lambda \over \mbox{TeV}}\right)^4\,,
\nonr\\
\left|C_T^{33,\tau \mu }\right|^2 + \left|C_T^{33,\mu \tau}\right|^2 <
6.8 \times 10^{-8} \left( {\Lambda \over \mbox{TeV}}\right)^4\,,
\nonr\\
\left|C_T^{33,\mu e}\right|^2 + \left|C_T^{33,e\mu}\right|^2 <
1.7 \times 10^{-7} \left( {\Lambda \over \mbox{TeV}}\right)^4\,.
\eeqa
We see that the lepton flavor changing tensor coefficients are constrained to be very small unless $\Lambda > 100$ TeV.

By the same token  we obtain  the following modification to the
charged lepton anomalous magnetic moment
\beq
\triangle a_{l} =  -{m_t m_l \over \pi^2
\Lambda^2}\ln\frac{\Lambda^2}{m_t^2}Re C_T^{33,ll}\,.
\eeq
>From the current limits  $|\triangle a_e|<3.5\times 10^{-11}$
and $|\triangle a_\mu|< 254 \times 10^{-11}$ \cite{PDG} we deduce that
\beqa
\left|{\frak{Re}} C_T^{33, ee} \right| <1.1 \times 10^{-3} \left({\Lambda\over
\mbox{TeV}}\right)^2\,,\nonr\\
\left|{\frak{Re}} C_T^{33, \mu\mu} \right| <4.1 \times 10^{-4} \left({\Lambda\over \mbox{TeV}}\right)^2\,,
\eeqa
which are less stringent than Eq.(\ref{LFVT}) but still very strong.

\subsection{ Triple Spin-Momenta Correlations in $e^+e^-\ra t \bar{t}$}

We give as an example a physical quantity measurable in the  process
at high energy which can be related to low energy constraints by RGE's.
Consider at the TeV linear collider (LC) the process
$e^+(p_+) e^-(p_{-},s_e)\ra t(k_+,s_t) \bar{t}(k_{-})$, where the momenta and spins are specified.
One can construct a  T-odd quantity involving two independent spins and a momentum ; e.g.
 $\hat{p}_{-} \cdot (\hat{s}_t \times \hat{s}_e)$ where  $\hat{p}_-$  is  the unit momentum 3-vectors, and $\hat{s}_e$
is the unit 3-spin vector in the electron rest frame.
 We can relate this to
$d_e$.

Since the tensor coefficient $C_T^{33,ee}$ has been determined to be around or smaller
than $10^{-9}$  from the electron EDM, Eq.(\ref{eq:boundT}), we need not consider it.
Only the effective
scalar operators will be important.
The triple correlation is a result of the interference between the SM amplitudes of
photon and $Z$ exchange, and the scalar 4-Fermi interaction. To
match the chirality, either the mass or the spin of electron has to be involved.
Clearly,  a LC with polarized electron beam offers higher possibility of  probing CPV part  of the scalar
interaction beyond SM.

A straightforward calculation yields the CPV amplitude, normalized to the photon exchanging
amplitude,
\beqa
|{\cal M}|^2_{TO} = {\frak{Im}} C_{S2}^{33,ee}(\Lambda)\frac{s}{3\Lambda^2}
\left[\left(1+\frac{m_t^2}{s}\right)(\hat{p}_--\hat{p}_+)\cdot(\hat{s}_t \times
\hat{s}_e)\right.\nonr\\
\left.+\left(1+\frac{2m_t^2}{s}\right)(\hat{k}_+-\hat{k}_-)\cdot(\hat{s}_t \times \hat{s}_e)\right]
\eeqa
where $s=(p_{+}+p_{-})^2$ is the Mandelstam variable, and $\hat{s}_t$ is the unit spin vector of the t-quark in
its rest frame.

In arriving at the above expression, we have assumed that $\sqrt{s}\gg M_Z$ and $\sin^2\theta_W
\simeq 0.25$ for simplicity.
Putting in  all the RG running we have derived,
Eqs.(\ref{eq:RGEST2Lam},\ref{eq:RGECS2TW}),
we get the following relation
\beqa
{ {\frak{Im}} C_{S2}^{33,ee}(\Lambda) \over {\frak{Im}} C_{S2}^{33,ee}(m_p) } &=&   (1.72)^{-1}\times(0.9456)\times
\left(\frac{m_t}{\Lambda}\right)^{0.036} \left({\Lambda \over
\mbox{TeV}}\right)^2\nonr\\
&=& 0.516\times\left({\Lambda \over \mbox{TeV}}\right)^{1.964}\,.
\eeqa
With the electron EDM constraint, Eq.(\ref{eq:CSboundEDM}), we conclude the
relative size of CPV amplitude has to be
\beq
< 2.7 \times 10^{-7}\times \left({\sqrt{s} \over \mbox{TeV}}\right)^2
\times \left({ \mbox{TeV}\over \Lambda}\right)^{0.036}\,.
\eeq
It is very challenging if not impossible to achieve this kind of precision at the ILC.

\section{Conclusions}
 We begin with the assumption that the SM gauge symmetry is valid from the some unknown new physics
scale $\Lambda$ to the EW scale and the matter content of the SM.
Using the effective field theory approach we studied all the dimension six operators which
contain a pair each of lepton and quark fields.
A previously derived result \cite{CN1} allows us to cast them in the mass eigenstates.
By a general assumption that the vector Wilson coefficients are constants we can eliminate zero order FCNC.
This leaves the most the scalar and tensor operators as the most interesting ones that can have FCNC at
the level of the Born term.
They also provide new sources of CP violation in both charge and neutral current channels. Furthermore,
in the NC channels there can be flavor changing as well as flavor conserving modes. For notational simplicity
we generically call the corresponding coefficients $C_S$ and $C_T$.
We also argue that stringent experimental limits make the Wilson coefficients of dimension six operator
of the dipole type negligible and permit us to concentrate on 4-Fermi operators.

An  important aspect of field theory is that these Wilson coefficients evolved with energy and this
can be calculated by the RG method. Once a coefficient is measured at one scale, say the EW scale, its
value at higher energy is determined by the anomalous dimensions. This robust prediction depends only on
the assumed gauge symmetry and the known states between the two energies. We have calculated the
anomalous dimensions of the complete set of semileptonic 4-Fermi operators to 1-loop. The RG running
of the associate Wilson coefficients are solved between the EW scale and $\Lambda$.

Below the EW scale the conserving gauge symmetries are $U(1)_{em}$ and color $SU(3)_c$. We found that the
effect due to $U(1)_{em}$ is not large in agreement with previous studies; however the QCD
corrections to $C_S$ and $C_T$ are large. The combine effect increases $C_S$ by almost a factor of
two from the scale of  $M_W$ to 1 GeV; whereas $C_T$ is decreased by approximately seventeen percent.

As is well known that pure leptonic meson decays are the most sensitive tests of
$C_S$,  we refined the previous analysis on the constraints by including RG effects. We also found that
the large logarithms of the RG method change the result on $C_T$ by almost a factor of two.
It turns out that the most stringent constraint on all the different $C_T$'s arise from 1-loop effects
they induce in EDM, anomalous magnetic moments, and rare decays of leptons.

The CPV effects from ${\frak{Im}}C_S$ do not give rise to EDM's at 1-loop. At 2-loops they
constraint the flavor diagonal operators; especially those involving the t-quark. Since these are
suppressed by an extra loop factor the bounds are not as tight as those for $C_T$. The flavor
off-diagonal ${\frak {Im}C_S}$ are most directly constrained by polarization measurements in semileptonic
meson decays such as in $K^+_{\mu3}$. Hence, these are very complementary tests in these very general
framework of effective operator approach to new physics. Furthermore, given the limits obtained at low
energies and the RG effects are not large we do not expect any of the $C_S$ coefficients to  be measurable
 at the LHC. On the other hand high energy colliders may be more
sensitive to the vector terms in ${\lag}_6^{\mathrm {NC}}$. This warrants a detail study which we
shall reserve for future investigations.

Theoretically the bounds we obtained on $C_S$ and $C_T$ have profound implications for model building.
They imply that if the new physics scale characterized by $\Lambda$   were to be below 10 TeV
any tree-level origin of these effective operators  must have small couplings. Although it is not impossible this
implies some degree of fine tuning is at work.  Well known examples are the multiple Higgs models
\cite{BG} and R-parity violating supersymmetric models. An alternative and perhaps more natural one  will
be that they are dynamically suppressed. For example they are generated at 1-loop or higher level
in the new physics model. Many supersymmetric models have this feature. Another possibility  will be
that the suppression is a result of some symmetry at work above $\Lambda$. Family symmetry is an
example that comes to mind which will suppress off diagonal terms and leave the diagonal ones relatively free.

We are looking forward to more precision measurements and hope that some of these operators will be measured.
Our analysis will be useful to unraveling the new physics at work in a general way.

\acknowledgments{J.N.N. would like to thank the Natural Science and Engineering Research Council of Canada for its support of
this work. The research of W.F.C. is supported by the
Academia Sinica Postdoctoral Fellowship, Taiwan.
}

\end{document}